\def\year{2020}\relax
\documentclass[letterpaper]{article} 
\usepackage{aaai20}  
\usepackage{times}  
\usepackage{helvet} 
\usepackage{courier}  
\usepackage[hyphens]{url}  
\usepackage{graphicx} 
\usepackage{graphicx}  
\usepackage{amssymb}
\usepackage{mathrsfs}
\usepackage{multirow}
\usepackage{amsmath}

\usepackage{color}

\title{Continuous Multiagent Control using Collective Behavior Entropy for Large-Scale Home Energy Management}
\author{Jianwen Sun$^{2,\dagger}$, Yan Zheng$^{1,2,\dagger}$, Jianye Hao$^{1,3,}$\thanks{Corresponding author. $\dagger$ Equal contribution.}, Zhaopeng Meng$^1$, Yang Liu$^2$\\
\textsuperscript{\rm 1}College of Intelligence and Computing, Tianjin University, Tianjin, China\\ 
\textsuperscript{\rm 2} Nanyang Technological University, Singapore\\
\textsuperscript{\rm 3} Noah’s Ark Lab, Huawei\\
\{yanzheng,jianye.hao,mengzp\}@tju.edu.cn, \{kevensun, yangliu\}@ntu.edu.sg
}
\begin{document}

\maketitle

\begin{abstract}
With the increasing popularity of electric vehicles, distributed energy generation and storage facilities in smart grid systems, an efficient Demand-Side Management (DSM) is urgent for energy savings and peak loads reduction.
Traditional DSM works focusing on optimizing the energy activities for a single household can not scale up to large-scale home energy management problems.
Multi-agent Deep Reinforcement Learning (MA-DRL) shows a potential way to solve the problem of scalability, where modern homes interact together to reduce energy consumers consumption while striking a balance between energy cost and peak loads reduction.
However, it is difficult to solve such an environment with the non-stationarity, and existing MA-DRL approaches cannot effectively give incentives for expected group behavior.
In this paper, we propose a collective MA-DRL algorithm with continuous action space to provide fine-grained control on a large scale microgrid.
To mitigate the non-stationarity of the microgrid environment, a novel predictive model is proposed to measure the collective market behavior. Besides, a collective behavior entropy is introduced to reduce the high peak loads incurred by the collective behaviors of all householders in the smart grid.
Empirical results show that our approach significantly outperforms the state-of-the-art methods regarding power cost reduction and daily peak loads optimization.
\end{abstract}

\section{Introduction}
Due to the growing number of various electric devices in modern life, meeting the energy demand becomes a significant challenge for the power grid ~\cite{wu2014neighborhood}. 
Companies spend tremendous time and money to satisfy the sharply changing demands.
With a fast-growing market share, the large-scale use of electric vehicles brings a high power consumption burden on the power system.
As more and more electric drive products are introduced into people's daily lives, the power grid is facing a major challenge of a large number of unstable loads.
Meanwhile, distributed renewable power generation, such as wind and solar energy, is considered to be vital to achieving cost and carbon reduction goals and is gaining prominence ~\cite{jain2014multiarmed}.
However, renewable power generation relies on unstable natural resources, which brings about a significant challenge to maintain demand-supply balance when consuming electricity from renewable sources. Meantime, with the rapid development of the advanced meters and basic computing facilities, smart grids with distributed energy generation and distributed energy storage have shown a high potential for energy management. Therefore, efficient energy-saving, management, and maximizing energy efficiency draw more and more attention.

Deep reinforcement learning (DRL) has achieved tremendous success not only in various of single-agent tasks~\cite{Silver2016MasteringTG,levine2016end,zheng2019diverse,zheng2019wuji}, but also complex multiagent scenarios~\cite{zheng2018deep,zheng2018weighted}. Recently, applying DRL in the smart grid has been studied~\cite{yan2018data,huang2019adaptive,duan2019deep}. \citeauthor{berlink2015batch} applied a RL-based demand-side management (DSM) technique to learn an optimal management policy in the energy management system (EMS)~\cite{berlink2015batch}. \citeauthor{di2018optimizing} studied the RL-based charing policy for electric vehicles (EV)~\cite{di2018optimizing}. \citeauthor{franccois2016deep} designed a DQN-based storage management strategy for a microgrid ~\cite{franccois2016deep}. However, these works investigate energy policies for a single household, restricting the algorithm's effectiveness and generalization in real scenarios.

From a multiagent system perspective, \citeauthor{dusparic2013multi} leverages the independent Q-learning to manage the energy demand of a small group of households, where each agent scheduled for the appliance usage of one household~\cite{dusparic2013multi}. However, their approach considered the role of the household as the traditional power consumer, and the rigid appliance usage schedules are inconvenient for flexible usage. 
The most related work to ours is that of \citeauthor{ijcai2019-89}, who focuses on a microgrid scenario and optimizes the energy expense problem with a multiagent reinforcement learning framework. They adopted the collective behavior approximations to further reduce the cost compared with the single household case and use the entropy-based method to reduce the high peak load~\cite{ijcai2019-89}.
However, the entropy-based rewards encourage the agents to take unpredictable actions such that EV cannot meet its charging objective during the low-price period. Moreover, same as other works ~\cite{berlink2015batch,di2018optimizing,dusparic2013multi}, \citeauthor{ijcai2019-89} only considers using discrete control actions by simply discretizing the continuous control values, which may encounter dimensionality issues with a large number of continuous action spaces.
The discretization unnecessarily throws away information about the structure of the action domain, which may be essential for precise control of energy management.

To address these, we extend the collective MA-DRL algorithm to continuous action space, enabling more flexible and fine-grained control. Besides, a predictive model is introduced to mitigate the non-stationarity by approximating the market behavior. Lastly, a collective behavior entropy is introduced to balance between maximizing single household's interests and encouraging moderate collective behavior for a smart home community.
Empirical results show that, compared with the state-of-art algorithms, our method achieves better reduction in terms of both costs and peak loads.

\section{Preliminaries}
\subsection{Markov Game}
Markov game, as an multiagent extension of Markov decision process (MDP), provide a commonly used framework for modeling interactions among agents. It can be formalized as an tuple $<N,S, \textbf{A},T,R,\lambda>$. Here $N$ is the
number of agents, $S$ is a set of states, $\textbf{A} = A_1\times \cdots \times A_N $ is the joint action set, where $A_i$ is the action space of agent $i$, $R$ is the reward function $S \times \textbf{A} \rightarrow \mathbb{R} $, $T $ is the transition function $ S\times \textbf{A} \times S \rightarrow [0,1]$. 
At each step $t$, agent $i$ receives its own observations $s_t$ and selects an action $a_t$ according to its policy $\pi_i(a_t|s_t)$. After every agent taking actions, the env transits into a new state $s_{t+1}$ according to the transition probability function $p(s_{t+1}|s_t, \textbf{a})$, where $\textbf{a}=(a_1,\cdots, a_i, \cdots, a_n)$ is the joint actions of all agents. Consequently, each agent $i$ receives immediate reward $r_i(s_t, \textbf{a})$ and new observation $s_{t+1}$. Then the process repeats. The goal is for each agent $i$ to find a policy $\pi_i$ that maximizes its own expected future reward 
$\mathbb{E}_{{a_i}\sim\pi_i} \sum\limits_{t=0}^\infty \lambda^t r_i(s_t,\textbf{a}),$
where $\lambda$ is a discount factor.

\subsection{Proximal Policy Optimization (PPO)}
Policy-based RL algorithms try to maximize the expected return $J(\pi_\theta)$ by leveraging the gradient with respect to the parameters of its policy $\pi_\theta$ as follows:
\begin{equation}
    \nabla_\theta J(\pi_\theta) = \mathbb{E}_{a\sim \pi_\theta}[\nabla_\theta \log(\pi_\theta(a_t|s_t))\hat{A}(s_t, a_t)]
\end{equation}
where $\hat{A}(s_t, a_t)$ is an estimator of the advantage function at timestep $t$. However, the estimation may vary dramatically between different runs, resulting in high variance. To address this, PPO measures a probability ratio between old and new policies $r(\theta) = \frac{\pi_\theta(a|s)}{\pi_{\theta_{old}}(a|s)}$, and imposes the constraint by forcing the ratio to stay within a small interval $[1-\varepsilon , 1+\varepsilon]$ as follows:
\begin{equation}
    \min \left(r(\theta) \hat{A}(s, a), \operatorname{clip}(r(\theta), 1-\epsilon, 1+\epsilon) \hat{A}(s, a)\right)
\end{equation}
where $\varepsilon$ is a hyperparameter that controls the clipping degree.

\section{Problem Formulation}
Following the concept in ~\cite{di2018optimizing,ijcai2019-89}, we illustrate the main components in the smart home as follows:
\begin{itemize}
  \item base load power consumption from conventional household appliances.
  \item power consumption for EV charging.
  \item micro-generation for renewable energy.
  \item home battery for electricity storage.
  \item home energy management system (EMS) for DSM.
\end{itemize}

\subsection{Microgrid Market}
Microgrid market consists of a great number of households, which are modeled as the basic unit to schedule their individual power trading plan. For households, price signals at different time slots are provided by the time-of-use (TOU) rate. At the beginning of every time slot, households need to make energy decisions (e.g., power trading or EV charging) based on the given TOU price, energy demand, power generation status, and other useful signals. The microgrid market can be modeled as a partially observable Markov Game, and below we state each component.

\subsubsection{States}
At time $t$, the observation (state) $s^i_t$ for household $i$ is formalized as $(p_t, H^i_{l,t}, H^i_{b,t}, H^i_{p,t}, E^i_{a,t}, E^i_{b,t}, E^i_{d,t})$, consisting of following components: 
\begin{itemize}
    \item $p_t$: electricity price at time $t$.
    \item $H_{l,t}$: power consumption of the base load.
    \item $H_{b,t}$: home-based battery state of charge. 
    \item $H_{p,t}$: energy amount of the home-based PV generation.
    \item $E_{a,t}$: EV charging availability. 
    \item $E_{b,t}$: EV-based battery state of charge.
    \item $E_{d,t}$: EV departure time.
\end{itemize}
Note that, $H_{b,t}$ and $E_{b,t}$ are 100\% when the battery is fully charged and 0\% when fully discharged. $E_{a,t}$ is set to 1 if EV is available and 0 otherwise. $E_{d,t}$ defines how many hours remains before EV departure.

\subsubsection{Actions}
Home EMS interacts with the power grid by adjusting the charging or discharging rate of EV and home-based batteries with a continuous rate. Thus, for each type of battery, every household has two alternatives: purchasing power from the grid (charging) and selling power to the grid (discharging). 

Formally, at each time slot $t$, home EMS chooses to decide two actions: power trading amount $P_{c,t}$ and the EV charging rate $C_{e,t}$.
For trading behavior, the value of $P_{c,t}$ is between [-$\delta$, $\alpha$], where $\delta$ and $\alpha$ are the average gross power generation and consumption per hour on the previous day, respectively. As for EV charging rate, following~\cite{di2018optimizing}, $C_{e,t}$ lies between [-$\eta$,$\eta$], where -$\eta$ and $\eta$ mean 100\% discharging and charging, respectively. Note that, for both behaviors, value $0$ means no trading or charging.
It is worth mentioning that, at each time slot $t$, the home battery is responsible for redundancy caused by $P_{c,t}$ and $C_{e,t}$ with satisfying the equation as follow:
\begin{equation}
  H_{p,t} + P_{c,t} = C_{e,t} + C_{b,t} + H_{l,t} 
\end{equation}
where $C_{b,t}$ is the charging rate of the home-based battery, and $C_{e,t}$ will be 0 if the EV is not available.
After deciding $P_{c,t}$ and $C_{e,t}$, the power surplus or shortage $C_t$ are measured as follows:
\begin{equation}
    C_t=H_{p,t}+H_{b,t}*B_h-H_{l,t}-C_{e,t},
\end{equation}
where $B_h$ is home battery capacity. If $C_t\neq0$ after performing $C_{e,t}$ and $P_{c,t}$, the home battery will be charged if power surplus or discharged if shortage. The battery charging rate and discharging rate are no more than the maximal charging rate while deciding the allowed actions. It is worth mentioning that, in our setting, EV charge capacity must reach more than 90\% of normal capacity before its departure time.

\subsubsection{Rewards}
One of our goals is to minimize household electricity costs.
We directly use the electricity cost as the incentive signal for the DRL agents.
The reward function is calculated as follows:
\begin{equation}
r_{t}=\left\{
             \begin{array}{lr}
             - P_{c,t}*p_b & if \ P_{c,t} \geq 0  \\
             - P_{c,t}*p_s & if \ P_{c,t} < 0 
             \end{array}
\right.
\end{equation}
where $p_b$ and $p_s$ are buying and selling price at current time slot, respectively.
The negative sign indicates that our goal is to minimize the overall energy cost.
Note that $p_b$ is always higher than $p_s$ due to the infrastructure cost of the power grid and the long-distance transmission fees.

\subsubsection{Transitions} At each time slot, all households make action decisions at first. Next, microgrid processes them and deterministically proceeds into the next state. 
After performing power actions, each household updates its state based on its energy management actions and TOU price.

\subsection{Group Incentive Mechanism}

We follow the collective microgrid model from ~\cite{ijcai2019-89}.
In the large-scale microgrid, each household can independently interact with the power grid according to its electricity demand.
However, these households are more willing to ally in order to coordinate the power production and consumption in a smaller community. 
This cooperation not only helps them balance the demand and supply in the power grid but also saves expenses for households.
Therefore, we devise an incentive-driven market mechanism to appeal agents to join a community for cooperation.
There are two trading processes in the microgrid market: the internal trading process and the external trading process.
At first, households trade inside the community and try to satisfy the demand inside.
If the internal trading inside the community fails to fulfill the needs, the external microgrid will trade with households and address the remaining demand.
The prices for internal trading and external trading are defined as follows:

$$p_{os,t} \leq p_{in,t} \leq p_{ob,t}$$

where $p_{os,t}$ and $p_{ob,t}$ are power selling and buying prices for external trading.
$p_{in,t}$ is the internal trading price.
In such a setting, each household prefers internal trading because it is cost-effective.
As for the entire microgrid, the internal optimization inside the group avoids some unnecessary inter-group trading, which is also beneficial for reducing the overall operating cost as well as peak load.
The final cleaning price for electricity integrated with external trading is:

\begin{equation}
 \begin{split}
p_{s,t}=\left\{
             \begin{array}{lr}
             \frac{p_{in,t}\varPsi_{b,t}+p_{os,t}(\varPsi_{s,t}-\varPsi_{b,t})}{\varPsi_{s,t}}, & \text { if } \ \varPsi_{s,t} \geq \varPsi_{b,t}  \\
             p_{in,t}, & \text { if } \ \varPsi_{s,t} < \varPsi_{b,t}
             \end{array}
\right.\\
p_{b,t}=\left\{
             \begin{array}{lr}
             p_{in,t}, & \text { if } \ \varPsi_{s,t} \geq \varPsi_{b,t} \\
             \frac{p_{in,t}\varPsi_{s,t}+p_{ob,t}(\varPsi_{b,t}-\varPsi_{s,t})}{\varPsi_{b,t}}, & \text { if } \ \varPsi_{s,t} < \varPsi_{b,t}
             \end{array}
\right.
\end{split}
\end{equation}
where $p_{s,t}$ and $p_{b,t}$ are current selling and buying price for external trading. 
$\varPsi_{s,t}$ and $\varPsi_{b,t}$ denote the total amount of selling and buying by the community.
The above settings ensure that $p_{s,t}$ and $p_{b,t}$ are always in [$p_{os,t}$, $p_{in,t}$] and $[p_{in,t} , p_{ob,t}]$, respectively.

In the large-scale microgrid, each household must consider the market dynamics to determine its trading strategy, and the market dynamics are affected by the actions taken by other households.
Thus, we turn the problem into a multiagent scenario where the community can be modeled as a multiagent system.
In order to better optimize costs, it is necessary to promote group coordination.
The problem is inherently multiagent and can be solved by MARL approaches.

\section{Algorithm}
In this section, we introduce the collective PPO algorithm with continuous action space to provide a more fine-grained control on a large scale microgrid.
To mitigate the non-stationary issue of the microgrid environment, we propose a novel supervised predictive model to simulate a joint representation of household actions, which is tightly adopted in our training process.
To solve the problem of excessively high peak loads of microgrid induced by the centralized distribution of charging behaviors, we reshape the reward function of our algorithm with an extra collective behavior entropy to reduce the loads without incurring additional costs.

\subsection{From Discrete to Continuous Control}
Like most real-world control tasks, home energy management is also a typical continuous control problem, as the EV battery and home battery can be charged or discharged at continuous rates.
Continuous action spaces are generally more challenging in reinforcement learning solutions~\cite{lillicrap2015continuous}.
\citeauthor{ijcai2019-89} leveraged discretization over continuous action space to bypass this challenge ~\cite{ijcai2019-89}.
However, discrete action space is impractical for a smart grid as the number of actions increases exponentially with the number of controlled values.
Furthermore, the situation is even worse for EMS as it demands precise control actions and more fine-grained discretization. 
It is challenging and time-consuming to explore efficiently in such a vast action space.
Moreover, the discrete action space discards some critical parts of continuous values.
For instance, in ~\cite{ijcai2019-89} where EV charging actions are configured as [-100\%, -50\%, 0, 50\%, 100\%] , it is impossible for one household to charge its EV at 25\% of the maximum charging rate.

To support continuous control, we deploy a Gaussian policy, which is defined as
\begin{equation}
  \pi_\theta(a|s)=\frac{1}{\sqrt{2\pi\sigma}}\exp(-\frac{(a-\mu)^2}{2\sigma^2}),
\end{equation}
where the mean $\mu=\mu_\theta(s)$ and standard deviation $\sigma=\sigma_\theta(s)$ are the outputs of the policy network parameterized by $\theta$.
To enable the use of back propagation, we adopt the idea of reparameterization ~\cite{heess2015learning} to write the $\pi(a|s)$ as a function $y=\mu_\theta(s)+\xi\sigma_\theta(s)$, where $\xi\sim\mathcal{N}(0,1)$.
The policy gradient with respect to $\mu$ and $\sigma$ can be computed as $\nabla_\mu \log \pi_\theta(a|s)=\frac{a-\mu}{\sigma^2}$ and $\nabla_\sigma \log \pi_\theta(a|s)=\frac{(a-\mu)^2}{\sigma^3}-\frac{1}{\sigma}$.

\subsection{Modeling Market Behavior}
In the microgrid market, hundreds of agents simultaneously decide each other's charging rate following its RL-based policies.
One issue in such an environment is that, from the view of any individual agent, the environment becomes non-stationary caused by the constant changes of other agents' policies during the training process.
One way to tackle this issue is centralized training with decentralized execution ~\cite{lowe2017multi}, where the critic can observe joint actions to approximate the value function.
However, as the dimension of the joint action space grows exponentially with the number of agents, it is difficult to learn the value function.

Different from traditional multiagent problems, each agent in a smart gird does not distributively interact with other agents to trade electricity.
Instead, each agent can directly interact with the centralized power grid.
Therefore, in our situation, the joint action of other agents could be abstracted as the microgrid market behavior. Consequently, the individual policy for agents $i$ can be formalized as:
\begin{equation}
\pi^i(s_t, a^{market}) \equiv \pi^i(s_t,a^1_t, ...,a^{i-1}_t, a^{i+1}_t, ..., a^N_t),
\end{equation}
where the market behaviors $a^{market}$ can be approximately measured by the seller group collective action $a_s$, the buyer group collective action $a_b$ and the group EV charging action distribution $\vec{C}_e$, changing the policy $\pi^i$ as follows:
\begin{equation}
\pi^i(s_t, a^{market}_t) \approx \pi^i(s^i_t, (a_{s,t}, a_{b,t}, \vec{C}_{e,t}))
\end{equation}
As a result, each agent can make decisions based on the market behaviors $a^{market}$ without knowing the specific behaviors of the others.

However, the market behaviors can only be garnered after all agents have made their decisions, resulting in a deadlock of dependencies.
\citeauthor{ijcai2019-89} tackled this issue by using group collective actions at the same time slot on the previous day to approximate current market dynamics~\cite{ijcai2019-89}.
It assumes there exist high similarities in market behaviors between consecutive days. 
However, this approach only considered daily similarities within 24 hours without taking into account the gap of the dynamic characteristics of the microgrid between a longer period, which can result in a poor approximation performance if the market dynamics change dramatically. 

To characterize the market behaviors more accurately, we propose a predictive model $P_\theta$ (parameterized by $\theta$) to directly predict the market behavior at time slot $t$ using the historical information of the microgrid as follows:
\begin{equation}
    \hat{a}^{market}_t = P(s_t, (s_{(t-n:t-1)},a^{market}_{(t-n:t-1)})), 
\end{equation}
where $s_t$, $s_{(t-n:t-1)}$ and $a^{market}_{(t-n:t-1)}$ denote the current observation, $n$ previous observations and market behaviors, respectively. The predictive model $P$ is trained using supervised learning. Based on more accurate prediction $\hat{a}^{market}_t$, agent $i$ can take action according to its policy at time slot $t$ as follows:
\begin{equation}
\pi^i(s_t, \hat{a}^{market}_t) = \pi^i(s^i_t, (a^p_{s,t}, a^p_{b,t}, \vec{C}^p_{e,t})),
\end{equation}
where $(a^p_{s,t}, a^p_{b,t}, \vec{C}^p_{e,t})$ are components of the predicted market behavior.

\subsection{Collective Behavior Entropy}

According to ~\cite{zhang2011fuel}, the average parking duration of EVs at night is more than 10 hours, and the immediate charging at home significantly surges and reach grid peak loads.
Despite RL-based learning algorithms could efficiently shift EV loads to the low-price period to reduce its operating cost, the uncoordinated EV charging behavior can still lead to new peak loads.
This phenomenon is caused by the selfish nature of the agents that every household trends to turn on the maximum EV charging rate when the low-price signal triggers.
Therefore, the smart grid DSM pursues a solution where every household can make full use of the off-peak period (about 8 hours) and charge their EVs at proper charging rates, rather than charging at a 100\% rate within 2 or 3 hours.

Individual entropy is proposed to diversify the EV charging behavior, which encourages the individual agent to take unpredictable actions by maximizing individual entropy~\cite{ijcai2019-89}. However, the charging efficiency remains low, as the EVs, to achieve diverse behavior, choose to discharge during the low-price period, resulting in poor performance. 
Instead of avoiding all kinds of collective behaviors, we only punish on the one that may lead to aggressively collective behavior (e.g., collective charging at maximum charging rates).

To achieve this, we propose the Collective Behavior Entropy (CBE), measuring from both the collective and individual perspectives as follows:
\begin{equation}
\label{reshaped_reward}
E_{cbe}(\pi^i) = D_{KL}(\pi^{col}||\pi^i) * D_{KL}(\pi^{ext}||\pi^i),
\end{equation}
where the KL-Divergence~\cite{kullback1951information} are employed to measure the similarity between policies. The former one measures the similarity between individual $\pi^i$ and collective policy $\pi^{col}$, describing the contribution of policy $\pi^i$ to the collective behavior. Note that, the collective policy is measured based on all actions in the microgrid (i.e., $\hat{\mu}=\bar{a}$, $\hat{\sigma}^2=\sum^N_{i=1}(a^i-\bar{a})^2$). 
The second entropy is employed to avoid extreme behavior (e.g., charging with a maximum rate). Assume $\pi^{ext}$ is the extreme behavior to be avoid, the similarity between $\pi^i$ and $\pi^{ext}$ can be measured by KL-divergence $D_{KL}(\pi^{ext}||\pi^i_t)$. Consequently, the combinations help to inhibit only extremely collective behavior rather than all collective behaviors. 

In DRL, CBE is used as a reward shaping item as follows:
\begin{equation}
\label{reshaped_reward}
r(\pi^i)=-\frac{\beta}{E_{cbe}(\pi^i)},
\end{equation}
where $\beta$ is the coefficient to adjust the impact of CBE on the ultimate reward function. Note that, 
due to the similarity being inversely proportional to the value of the KL divergence, here we use the reciprocal of the CBE in reward shaping. Reward shaping item $r(\pi^i)$ will prevent householder $i$ from charging the EV using a maximum rate when the grid load is heavy.

Intuitively, CBE will not punish the phenomenon that all households charge EVs concurrently using a moderate charging rate (e.g., less than 30\%). Besides, it will also not punish the maximum charging behavior if only a few householders are charging. This guarantees the charging efficiency of EV while reducing the high peak loads.

\section{Experiments}
In this section, we evaluate our proposed approach on simulated microgrid built on real-world data. We first introduce our data configuration of the problem, baselines, hyperparameters of the model. Then comparisons of related baselines in terms of cost reduction and peak load reduction are conducted to evaluate the proposed algorithm. Detail analysis and discussions are given in each evaluation.

\subsection{Experimental Settings}
\subsubsection{Microgrid Setup} 
The microgrid environment in this paper is built on various types of real-world data.
A random sample of 200 household power consumption items comes from the Midwest region of the United States data set ~\cite{muratori2018impact}.
The time-of-use electricity price, EV configurations, and home battery are from ~\cite{di2018optimizing}.
We calculate the EV status when arriving home based on the daily driving distance data, which obeys gamma distribution ~\cite{lin2012estimation} with shape 1.6 and scale 20.
The charging efficiency for home battery and EV battery is set as 0.9.
We use the PV generation data from ~\cite{queensland}.

\subsubsection{Baselines} The methods that we evaluated include (i) Naive, (ii) DQN, (iii) MA2C, (iv) MA2C-EB, (v) PPO, (vi) MPPO, (vii) PRE-MPPO, (viii) PRE-MA2C, (ix) PRE-MPPO-CBE, (x) MA2C-CBE.
The Navie method is a rule-based control algorithm described in ~\cite{berlink2015batch}, which sells all energy surplus at each instant and is regarded as the worst-case baseline.
The DQN method is introduced by ~\cite{di2018optimizing} and performs best in a single household case.
MA2C stands for Multiagent Actor-Critic method from ~\cite{ijcai2019-89}, which is augmented from A2C with market dynamics approximation to represent the collective group behavior.
MA2C-EB is the entropy-based collective A2C in ~\cite{ijcai2019-89} using individual entropy to encourage diverse group behavior, which is the baseline for peak load reduction.
The next six methods the ones we proposed or optimized with our approaches.
PPO is a continuous control policy-gradient method using PPO to update policy parameters.
MPPO is the multiagent PPO method leveraging the same group behavior approximations as MA2C ~\cite{ijcai2019-89}.
PRE-MPPO and PRE-MA2C are extended with predicted market dynamics.
PRE-MPPO-CBE and MA2C-CBE are equipped with Collective Behavior Entropy as an extra reward shaping method to reduce peak loads.

\subsubsection{Hyperparameters}
The discount factor $\gamma$ for MPPO is 0.99, and the clip parameter $\epsilon$ of PPO is 0.2.
There are two dense layers for all neural networks and each with 64 neurons for all multi-agent models.
The output layer of the network in PPO is the Gaussian distribution (e.g., $\mu$ and $\sigma$) for continuous control.
The output actions are stochastic and sampled from these distributions.
The learning rate of policy and value network for PPO is set as 0.0003 and is decayed linearly with the learning episode.
We use the first-28-day data for training and the rest of the data for testing.
We train all the algorithms for 125 episodes.
In each episode, $N$ homogeneous agents determine their action $a^i$ parallelly to jointly optimize one shared policy.
All methods are implemented by Python and run on a server with 12-core Intel(R) Xeon(R) CPU E5-1650 v3 @ 3.50GHz processors and 1 NVIDIA GTX 1080Ti GPU.

We use the group daily operation cost of the training process and total operation cost in testing as the evaluation criteria for cost reduction.

\subsection{Evaluation on Cost Reduction}

\subsubsection{Continuous Control.} 
\begin{figure}[t]
  \centering
  \includegraphics[width=0.95\linewidth]{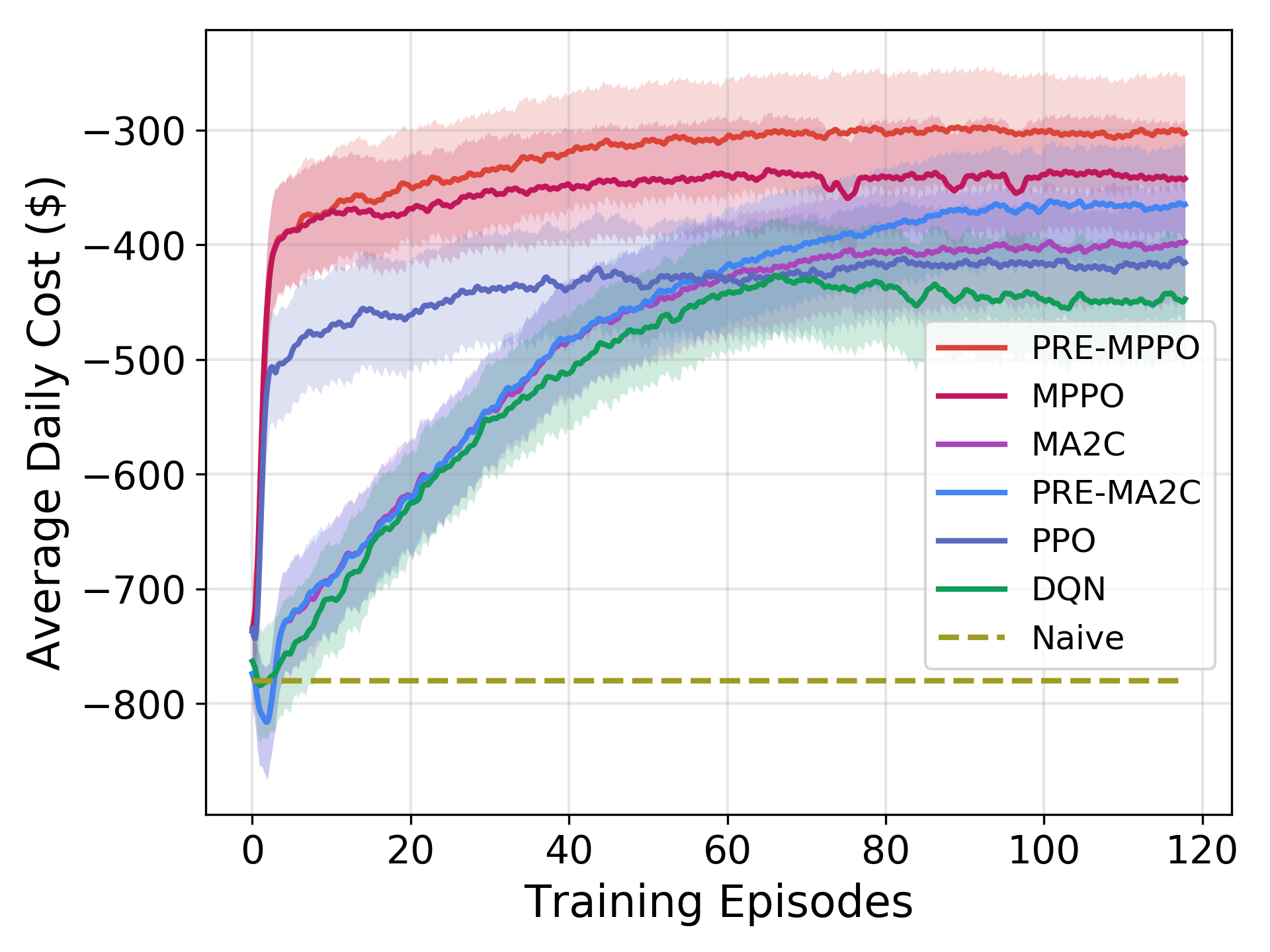}
  \caption{Comparisons regarding the average daily cost.}
\label{cost-compare}
\end{figure}

We first compare Naive, DQN, MA2C, PPO, and MPPO to validate the performance improvement of continuous action space.
PPO and MPPO are our proposed approaches using continuous action space, and the DQN and A2C are baselines with discrete action space.
In DQN and MA2C, the EV charging action and the trading action are configured as 5 and 9 discrete action, respectively, resulting in a combined 45-dimension action space.
As illustrated in Fig. \ref{cost-compare}, MPPO outperforms DQN and MA2C, both in the final performance and data efficiency.
Compared with MA2C, MPPO only uses 3\% of episodes to reach the same performance.
This result mainly comes from the adoption of continuous action space as exploring continuous action space is more efficient than high-dimensional discrete ones.

\begin{table}[t]
\centering
\begin{tabular}{crcc}
\hline
\hline 
\multirow{2}{*}{Algorithm} & \multirow{2}{*}{Operating Cost} & \multicolumn{2}{l}{Daily Peak Load}\tabularnewline
\cline{3-4} \cline{4-4} 
 &  & mean  & std \tabularnewline
\hline 
\hline 
Naive  & -263,195 (\$)  & 1360.10  & 53.27 \tabularnewline
DQN  & -113,674 (\$)  & 1249.22  & 28.87 \tabularnewline
MA2C  & -87,632 (\$)  & 1359.94  & 31.16 \tabularnewline
MA2C-EB  & -88,351 (\$) & 1194.83  & 34.80 \tabularnewline
PPO  & -97,683 (\$)  & 1375.20  & 45.81 \tabularnewline
MPPO  & -74,437 (\$)  & 1421.70  & 42.15 \tabularnewline
PRE-MPPO  & -67,372 (\$)  & 1432.58  & 43.59 \tabularnewline
PRE-MA2C  & -80,922 (\$)  & 1447.75  & 29.37 \tabularnewline
PRE-MPPO-CBE  & -74,559 (\$) & 876.28  & 115.71 \tabularnewline
MA2C-CBE  & -89,852 (\$) & 838.25  & 51.44 \tabularnewline
\hline
\hline
\end{tabular}
\caption{Comparisons of related baselinse in terms of the cost reduction.}
\label{table-cost}
\end{table}

Table \ref{table-cost} shows the operating results of the smart grid configured with the evaluation data.
All results are obtained as an average of 20 random seeds.
Naive method performs poorly as it charges EV once available, ignoring TOU prices and trades electricity on-demand without using electricity storage.
MPPO outperforms single household DQN by 34.5\% and collective MA2C by 15.1\%.
PPO achieves better cost reduction than DQN, but worse than MA2C.
This indicates that only using continuous control is not enough to achieve efficient optimization, and leveraging collective behavior is beneficial for achieving better control.

\subsubsection{Predictive Market Behavior.} To demonstrate the advantage of the adoption of predictive market behavior, we conduct experiments to compare the performance among PRE-MPPO, PRE-MA2C, MPPO, and MA2C.
As shown in Fig. \ref{cost-compare}, algorithms with predictive group behavior are superior to the ones with market dynamics approximations.
Table \ref{table-cost} shows that PRE-MPPO gains 9.5\% less operating cost compared with MPPO, while PRE-MA2C's cost is less than MA2C's by 7.7\%.
The reason is that predicted market behavior $(a^p_{s,t}, a^p_{b,t}, \vec{C}^p_{e,t})$ takes the periodicity of the grid as well as the dynamic changes of the market into account resulting in more realistic market behavior.
The results also demonstrate that algorithms can achieve higher performance with more accurate group behavior estimation. 

Another interesting finding is that only PPO, MPPO and PRE-MPPO perform a steep rise at the early stage, suggesting that continuous action space indeed contributes to achieving a flexible and accurate control.

\subsection{Evaluations on Peak Load Reduction}

\begin{figure}[t]
\centering
  \includegraphics[width=0.95\linewidth]{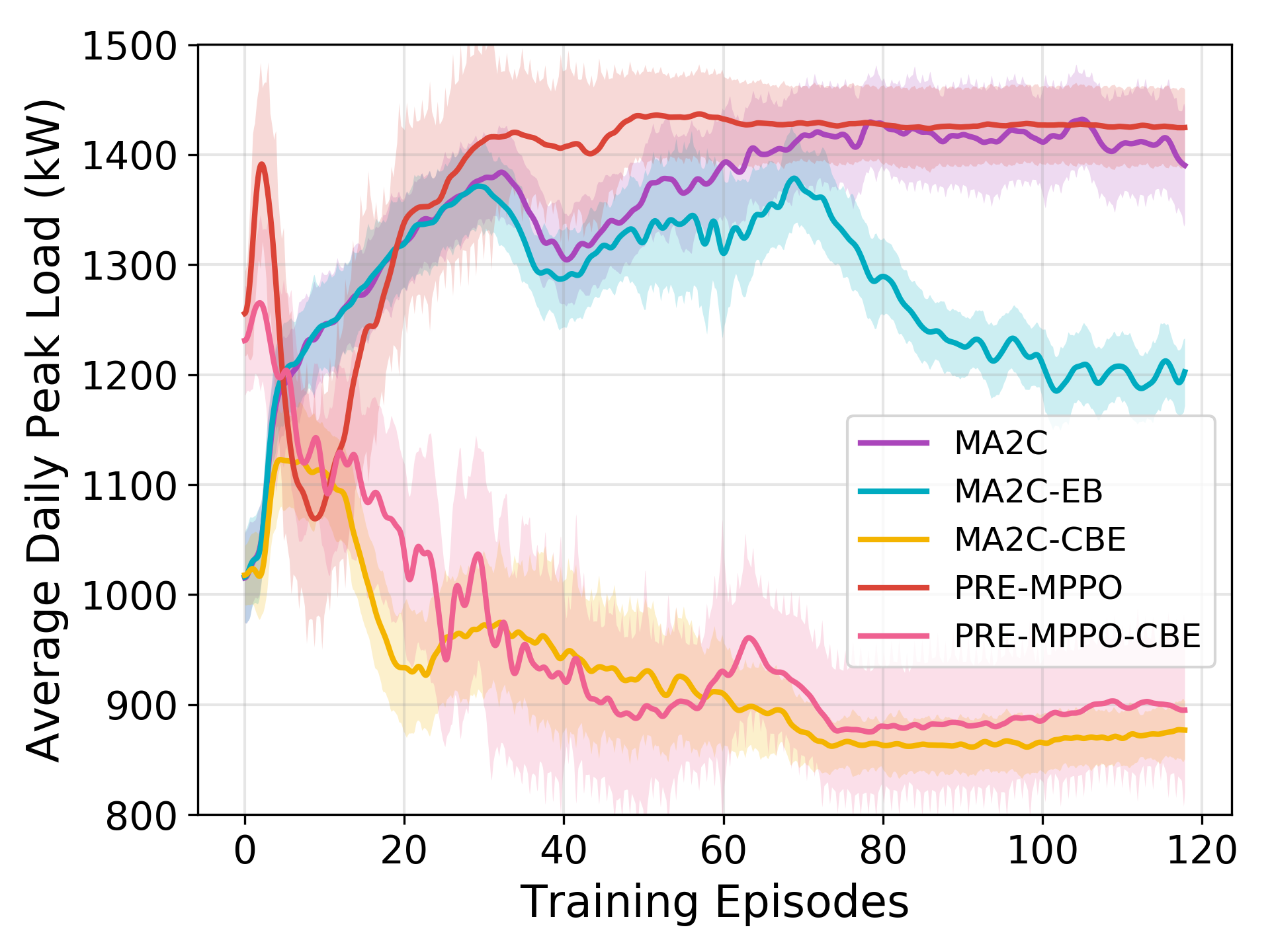}
  \caption{Comparisons regarding the average peak load.}
\label{load-compare}
\end{figure}

\begin{table}[t]
\begin{tabular}{crcc}
\hline
\hline
\multirow{2}{*}{Algorithm} & \multirow{2}{*}{Operating Cost} & \multicolumn{2}{l}{Daily Peak Load}\tabularnewline
\cline{3-4} \cline{4-4} 
 &  & mean  & std \tabularnewline
\hline
\hline
MA2C  & -87,632 (\$)  & 1359.94  & 31.16 \tabularnewline
MA2C-EB  & -88,351 (\$) & 1194.83  & 34.80 \tabularnewline
MA2C-CBE  & -89,852 (\$) & 838.25  & 51.44 \tabularnewline
PRE-MPPO  & -67,372 (\$) & 1432.58  & 43.59 \tabularnewline
PRE-MPPO-CBE  & -74,559 (\$) & 876.28  & 115.71 \tabularnewline
\hline
\hline
\end{tabular}
\caption{Comparisons regarding peak load reduction.}
\label{table-load}
\end{table}

We use the statistic data of the highest load in 24 hours of a day as the criterion to evaluate our method.
The high peak load data with lower variance means more stable performance.

\begin{figure}[t]
  \centering
  \includegraphics[width=0.95\linewidth]{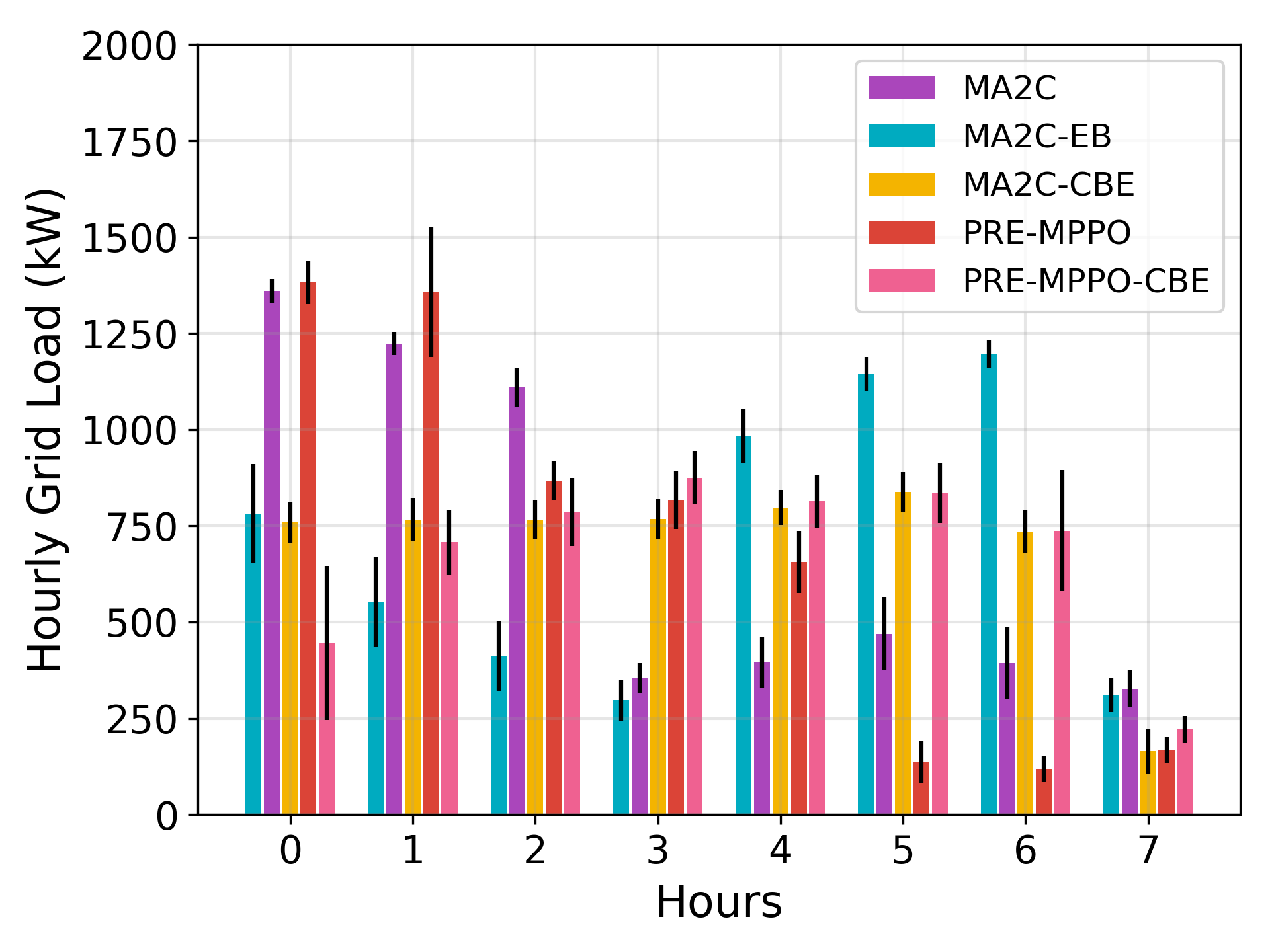}
  \caption{The average hourly peak load of related algorithms at night.}
\label{hourly-loads}
\end{figure}

As shown in Table \ref{table-cost}, since RL-based MPPO has achieved significant performance in reducing operating cost, its daily high peak load is even higher than the rule-based baseline Naive.
The reason is that cost-sensitive incentives encourage households to discharge EV at the high-price period and charge EV when the power price is low to save more energy costs.
The consequence is that all households choose to charge EV  when a low price signal triggers lock of coordination, thereby generating a new peak in smart grid load (0 am to 1 am for PRE-PPO, and 0 am to 2 am for MA2C shown in Fig. \ref{hourly-loads}).

As the peak load is the key indicator to measure grid performance, we conduct experiments to evaluate our method as well as the baseline and try to investigate their contribution to reducing the peak load.
The learning curves of various algorithms are shown in Fig.~\ref{load-compare}, and the evaluation results are shown in Table~\ref{load-compare}.

First, PRE-MPPO-CBE, compared with PRE-MPPO, achieves a 38.8\% peak load reduction at an additional 10\% electricity cost.
This indicates PRE-MPPO-CBE achieves a better balance by sacrificing some households’ benefits for the community.
By comparing MA2C-EB with MA2C, we can see that after 70 episodes of training, MA2C-EB starts to learn to reduce peak load and reaches 12.1\% lower peak load compared with MA2C.
Unlike MA2C-EB, with the help of group behavior entropy, MA2C-CBE starts to reduce peak loads at the early stage and finally reaches a 38.4\% peak load reduction compared with MA2C.

\subsection{Analysis of Charging Behavior \& Hyperparamters}
To explain the difference between normal methods (MA2C and PRE-MPPO), individual entropy (MA2C-EB from ~\cite{ijcai2019-89}), and our collective behavior entropy (MA2C-CBE and PRE-MPPO-CBE), we visualize the mean as well as the variance of hourly load in low price period (0 am to 8 am in our TOU data set) in Fig. \ref{hourly-loads}.
The results show that both PRE-MPPO and MA2C choose to charge EV from at high charging rates from 0 am to 1 am.
The charging rate is pretty low in the later period (5 am to 6 am for PRE-MPPO and 3 am to 6 am for MA2C), thereby they do not make full use of the low-price period.
The entropy-based MA2C-EB reduces the load from 0 am to 3 am. However, the loads from 4 am to 6 am are still high as in our environment settings, the EV must be charged to 90\% of the full power before departure time (7 am).
It is because the individual entropy encourages each household to take action different from others, and households may choose to discharge in the low-price period to get a higher bonus.
Therefore, the entropy-based approaches charge EVs insufficiently, resulting in another high peak (4 am to 6 am for MA2C-EB).
Different from MA2C-EB, the hourly load of our method (MA2C-CBE) is balanced during the low-price period, and the high peak is significantly lower than the entropy-based ones (MA2C-EB).
The reason is that our methods only punish aggressively collective charging and does not punish other collective behavior, e.g., moderate collective charging (0 am to 6 am for MA2C-CBE and PRE-MPPO-CBE).

\begin{figure}[t]
  \centering
  \includegraphics[width=0.95\linewidth]{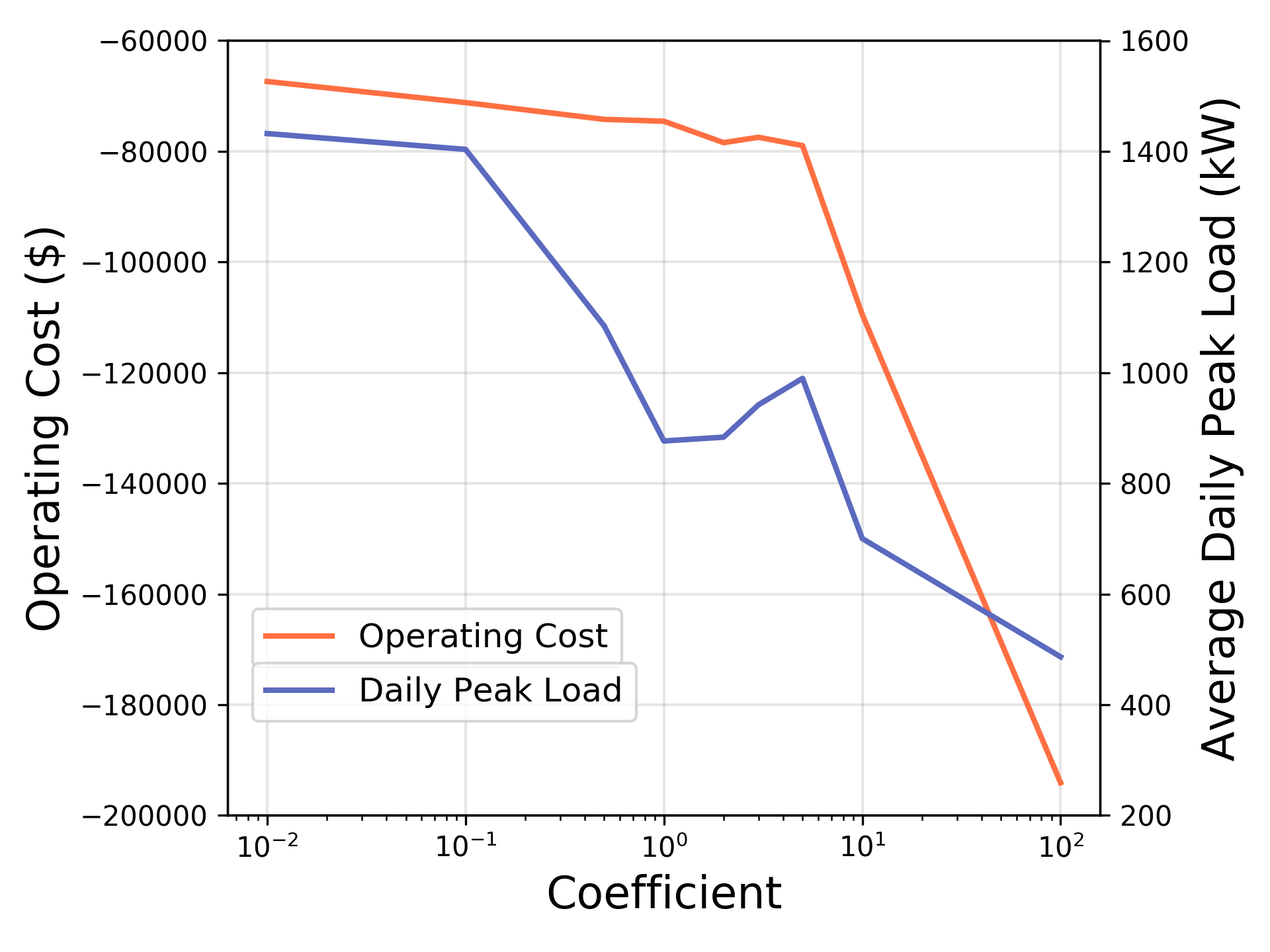}
  \caption{Grid operating results of PRE-MPPO-CBE with different coefficients.}
\label{coef-compare}
\end{figure}

To search for better coefficients for the reshaped reward function ($\beta$ in Equation \ref{reshaped_reward}), we use different coefficients to train our PRE-MPPO-CBE.
As shown in Fig. \ref{coef-compare}, PRE-MPPO-CBE achieves better peak load reduction by setting $coef$ to [1.0, 3.0].
With the $coef$ being larger than 10.0, the daily peak load could reach a low value of about 500 kW while the operating cost drops dramatically.
The reason is that the agents pay excessive attention to stay far away from aggressively collective charging without considering reducing the operating cost.

\section{Conclusions}
In this paper, we focus on the large-scale home EMS optimization problem for smart homes. A collective MA-DRL algorithm is proposed with continuous action space to achieve flexible and precise control on a large scale microgrid. Besides, a novel predictive model is proposed to mitigate non-stationarity in the microgrid environment by approximating the market behavior. Lastly, a collective behavior entropy is introduced to address the high peak loads incurred by the collective behaviors of the smart grid. Empirical results demonstrate that all these innovations altogether contribute to achieving better reduction in terms of both costs and peak loads than the state-of-the-art algorithms.

As future work, more effective models could be investigated to predict the microgrid market behavior accurately. In addition, our approach is designed from the community level to reduce the total cost and peak load for the overall community, given that all households are willing to follow the strategy learned from our framework. However, this situation may not be stable, and there may exist other strategies for an individual household to increase his utility (further reduce cost at the cost of increasing peak loads of the community). This is an interesting direction to further explore how to make the framework incentive-compatible.

\section{Acknowledge}
The work is supported by the National key R\&D program of China~(Grant No.: 2018YFB1701700), and the National Natural Science Foundation of China (Grant Nos.: 61702362, U1836214), 
the National Research Foundation, Prime Ministers Office Singapore under its National Cybersecurity R\&D Program (Award No. NRF2018NCR-NCR005-0001), 
National Satellite of Excellence in Trustworthy Software System (Award No.NRF2018NCR-NSOE003-0001) administered by the National Cybersecurity R\&D Directorate Singapore, 
and NTU research grant NGF-2019-06-024.

\bibliographystyle{aaai}
\bibliography{references}

\end{document}